\documentclass[conference]{IEEEtran}

\usepackage{graphicx}
\usepackage{cite,amsmath,amsfonts,graphicx}
\usepackage{bm}
\usepackage{amsbsy}
\usepackage{color}
\usepackage{soul}
\usepackage[caption=false]{subfig}

\begin{document}

\title{State Monitoring for Situational Awareness \\in Rural Microgrids Using the IoT Infrastructure}

\author{\IEEEauthorblockN{Seyed~Amir~Alavi,~\IEEEmembership{Graduate Student Member,~IEEE,}			
            Mehrnaz~Javadipour,~\IEEEmembership{Student Member,~IEEE,}
            Kamyar~Mehran~\IEEEmembership{Member,~IEEE}}
\IEEEauthorblockA{School of Electronic Engineering and Computer Science\\
Queen Mary University of London\\
London E1 4NS, UK\\
s.alavi@qmul.ac.uk}}

\maketitle

\begin{abstract}
This paper presents an event-triggered estimation strategy and a data collection architecture for situational awareness (SA) in microgrids. An estimation agent structure based on the event-triggered Kalman filter is proposed and implemented for state estimation layer of the SA using long range wide area network (LoRAWAN) protocol. A setup has been developed which can provide enormous data collection capabilities from smart meters, in order to realise an adequate SA level in microgrids. Thingsboard Internet of things (IoT) platform is used for the SA visualisation with a customised dashboard. It is shown by using the developed estimation strategy, an adequate level of SA can be achieved with a minimum installation and communication cost to have an accurate average state estimation of the microgrid.
\end{abstract}


\IEEEpeerreviewmaketitle

\section{Introduction}
Smart grid is the recently proposed structure for the modernized power system. The main characteristic of the proposed smart grid is better utilisation of renewable energy sources (RESs) and distributed generations (DGs), toward having the stabilized, reliable and resilient power system. This requirement with the complexity of information and communication technology (ICT) in smart grid, mandates the division of the power system into subsystems called "microgrids" \cite{Montazeri2013}. These subsystems operate autonomously in both islanded mode and connected mode regarding to the main grid. Microgrids operate differently regarding their applications and targets designed for, but they should be able to provide a common interface to the main grid to be integrated optimally.

Rural microgrids and smart villages are specific types of microgrids that pose unique requirements in terms of availability, cost, and operation \cite{Anderson2017}. The basic building blocks of "smartness" include access to high-quality education, healthcare, information and communication technologies, finance, and clean water and sanitation. However, underlying these building blocks lies one important element: energy \cite{Anderson2017}.

The main sources of energy accessible in rural areas are usually renewable energy sources. Photovoltaic (PV) panels and fuel cells (FCs) are among the cheapest available energy sources. These sources are intrinsically direct current (DC), that if integrated correctly, removes the need for costly DC-AC converters and decreases electrical loss \cite{Alavi2018a}. Also, DC loads such as mobile phones, digital assets and LEDs for lighting, form the main part of home appliances. As the main type of electrical loads in villages are home appliances, DC microgrids are proposed for the implementation of them recently \cite{Alavi2018a}.

Low cost installation and operation of these types of microgrids have been the topic of research for the last few years \cite{Sharifzadeh2017a}. With the introduction of the Internet of things (IoT) communication infrastructure, the cost of communications has been reduced to a great extent. Therefore, the trade-off between controller complexity and data communication cost has become simpler to reach a solution \cite{Alavi2016}.

In this paper, the authors have proposed a distributed event-based Kalman estimation filter for average consensus toward situational awareness of the rural microgrid. In Section \ref{sasg}, situational awareness domains related to microgrid are discussed and also the control and operation requirements are provided. In Section \ref{dacp}, the proposed estimation strategy analysis is provided with the developed event-based Kalman estimator. In Section \ref{prototype}, the implemented setup for evaluation of the esimation strategy is shown. Lastly, Section \ref{results} presents the results of the analysis. The paper is concluded in Section \ref{conc}.

\section{Situational Awareness-Centric Microgrids}\label{sasg}
Traditionally, the SCADA systems in microgrids monitor the basic variables and states to control both the power quality and demand response. One of the primary advantages of the microgrids is the enhanced grid monitoring, from the demand response and power quality to the smart user behaviour \cite{Basu2016}. This makes the smart grid not a sole entity providing services to consumers, but also a collection of different systems and technologies cooperating together to bring the highest reliability at the lowest cost with the participation of power consumers.

A smart grid is a multilayer, distributed, and multidomain system in which different types of operations are taking place in tandem \cite{Diao2010}. A close coordination should occur among all the players with different goals. Naturally, SA is designed for such a smart grid should be multilayered where in each layer, a team distributed SA handles the versatile tasks, such as the load forecasting, equipment health monitoring, power quality monitoring, cybersecurity, and unit commitment.

New functionalities of the modern grid comes with the cost of security risks of connection to the Internet. If in the past, industrial systems were considered secure, due to the use of proprietary controls and limited connectivity, smart grid increases the exposure of SCADA systems, and consequently sustaining the security issues in the network. Implementation of the adequate SA for micorgrids still needs to be developed further in a number of crucial areas, including \cite{Monajemi2017, Wang2014, Alavi2018}:

\begin{itemize}
\item \textit{Communication infrastructure monitoring} 
Faulty operation of the communication devices can lead to a verity of unknown issues bringing the entire system down. Hence, the efficient monitoring of the operation of the switches and advanced metering infrastructure (AMI) gateways would guarantee a reliable system.

\item \textit{Equipment health monitoring}
Health monitoring of the equipment especially in the distribution systems, is possible through the IoT platforms. Every equipment can be monitored from different control centres across the multilayered grid. The SA system can use health monitoring data in order to make pre-emptive decisions. For instance, number of failures in protection devices in a specific area of smart grid, can pinpoint the design problems in that specific area. This can be used for the prioritised maintenance.

\item \textit{Power generation and consumption}
The unit commitment in a smart grid is different from the one in a traditional grid, in which the power generation programming are merely based on the behaviour of the two players (i.e., power plant constrains, as well as load dynamics). Each player makes decisions based on different factors, such as weather and time. In a smart grid, number of players are continuously increasing due to different structure in the decision making. Distribution companies seek high income, where the consumers need cheap energy. The adequate SA can accommodate this multiplayer decision-making scenarios, in order to improve the power system operation and planning.

\item \textit{Microgrids connection status}
Different logics operate behind the microgrid operation which decides that in which situations the micorgrids should operate in an islanded mode or in a connected mode. Awareness of when and why a microgrid gets disconnected from the main grid makes the unit commitment and tertiary control cost effective and resilient to the faulty and abnormal situations.

\item \textit{Cybersecurity}
One of the most important factors in the adequate SA framework is the cyber SA. The appropriate analysis of the network traffic, as well as the distributed intrusion detection systems (DIDS) can be employed to increase the security of the overall SA-centric system.
\end{itemize}

The presented topics by no means are the exhaustive list of the research topics in an adequate SA system, but it covers the most important issues which can be used as a starting point. The common technologies available to villages in rural areas for power generation are solar panels, micro-wind electric, micro-hydro electric and biogas. In order to efficiently extract power for the sources, the DC-DC converters should operate in maximum power point tracking mode. Villages mandate several constraints in terms of control strategy and communication infrastructure, which are analysed in the following.

The main control requirements for smart villages are:
\begin{itemize}
\item The controller should be decentralized and distributed. As smart villages don't have a central authority to monitor microgrid online, distributed autonomous controllers provide reliable operation.
\item Distributed controllers should support Plug-and-Play (P\&P) operation. This is in fact a must for villages as they usually expand in time, so new sources of energy should be able to utilized easily without any interruption to the other users.
\item Excess power generated by DGs should be stored for times that the power is not available.
\end{itemize}


\section{Distributed Average Consensus Protocol}\label{dacp}
Each estimation agent has an average state estimator that uses the local measurements and information from the neighboring agents to update the local estimates of the average microgrid quantities. The average state estimator implements a distributed average consensus protocol for tracking the dynamic signals from \cite{Spanos2005DynamicCF} and \cite{Morstyn2016}.

The agents are connected by a sparse communication graph $\mathcal{G(V,E)}$ with the nodes $\mathcal{V}=(1,...,\mathcal{N)}$ and edges $\mathcal{E}$. Each graph node represents an estimation agent, and the graph edges represent communication links between them. $(i,j)\in \mathcal{E}$ if there is a link allowing information flow from node $i$ to node $j$. The neighbours of $i$ node are given by $\mathcal{N}_i$, where $j \in \mathcal{N}_i$ if $(j,i)\in \mathcal{E}$. The graph adjacency matrix is given by $A = [a_{ij}] \in R^{N \times N}$, where $a_{ij} > 0$ if $(j,i)\in \mathcal{E}$ and $a_{ij}=0$ otherwise.

For the $i$th ES system, let $x_i$ be a local state variable, and let $\overline{x}_i$ be a local estimate of the average value of that state for the ES systems. The $i$th estimation agent receives the average state estimates from its neighbours $j \in \mathcal{N}_i$, and its average state estimator implements the following distributed average consensus protocol:
\begin{equation}
\overline{x}_i=x_i+\int \sum\limits_{j \in \mathcal{N}_i}{a_{ij}(\overline{x}_j-\overline{x}_i)dt}
\end{equation}

Each node in the network has in-degree $d_i =\sum_{j=1}^{N} a_{ij}$ and out-degree $d_i^o = \sum_{j=1}^{N} a_{ji}$. Moreover, the graph is balanced if $d_i = d_i^o$ for all the nodes. The graph degree matrix is given by $\textbf{D} = diag\{d_i\}$ and the graph Laplacian matrix is also given by $\textbf{L}=\textbf{D}-\textbf{A}$. The global dynamics of the distributed average consensus protocol are given by:
\begin{equation}
\dot{\overline{\textbf{x}}} = \dot{\textbf{x}}-\textbf{L}\overline{\textbf{x}}
\end{equation}

Applying the Laplace transform yields the following transfer function matrix for the average consensus protocol \cite{Spanos2005DynamicCF}:
\begin{equation}
G^{avg}=\frac{\overline{\textbf{X}}}{\textbf{X}}=s(sI_N+\textbf{L})^{-1}
\end{equation}

$\overline{\textbf{X}}$ and $\textbf{X}$ are the Laplace transforms of $\overline{x}$ and $x$, respectively.

For a balanced communication graph with a spanning tree, the steady-state gain of the average consensus protocol is given by the averaging matrix \cite{DePersis2018}:
\begin{equation}
\lim_{s\to 0} G^{avg}= Q \mathrm{,where}\ [Q]_{ij}=\frac{1}{N}
\end{equation}
The final value theorem shows that for a vector of step inputs, the elements of $\overline{\textbf{x}}(t)$ converge to the global average of the steady-state values $\textbf{x}^{ss}$:
\begin{equation}\label{eq:dafafasaxz}
\lim_{t\to \infty} \overline{\textbf{x}}(t)= \lim_{s\to 0} G^{avg}\lim_{t\to \infty} s\textbf{X}=Q\textbf{x}^{ss}=\left\langle \textbf{x}^{ss} \right\rangle \underline{1} 
\end{equation}

\subsection{Event-Based Kalman Filter Design}
Consider the following linear system which is the state space realization of distributed average consensus protocol transfer function in each estimator agent:
\begin{IEEEeqnarray}{ll}
\dot{x}=Ax\left(t\right)+w\left(t\right)\IEEEnonumber \\ y(t)=Cx(t)+v(t)
\end{IEEEeqnarray}
where $x\mathrm{\ }\mathrm{\in}\mathrm{\ }{\mathrm{R}}^{\mathrm{n}}$ is the estimated state and $y\ \mathrm{\in}\mathrm{\ }{\mathrm{R}}^{\mathrm{p}}\mathrm{\ }$is the output measurement. The process noise $w\left(t\right)$ and measurement noise $v(t)$ are the uncorrelated, zero-mean white Gaussian random signals, fulfilling the following:
\begin{IEEEeqnarray}{lll}
E\left\{w(t) \; w(s)'\right\} = Q \; \delta (t-s)
\\
E\left\{v(t) \; v(s)'\right\} = R \; \delta (t-s)
\\
E\left\{w_i\left(t\right){v_j\left(s\right)}'\right\} =0,\;\;1 \leq i \leq n, \;\;1 \leq j \leq p
\end{IEEEeqnarray}
where $w_i$ and $v_j$ are the \textit{i}-th and \textit{j}-th elements of the $w$ and $v$, respectively. Also, $R$ is the measurement noise covariance, and $Q$ is the process noise covariance. It is assumed that the\textit{ i}-th sensor only transmits the data when the difference between the current sensor value and the previously transmitted value is greater than ${\delta }_i$.

The states are also estimated periodically with the period of $T$. For simplicity, it is assumed that there is no delay in the sensor data transmission. Using the SoD method \cite{Li2016}, the estimator continuously with a period of $T$ demands the data from the sensors no matter the data becomes available. For example, if the last received $i$-th sensor value is $y_i$ at the time $t_{last,i}$, and there is no $i$-th sensor data received for ${t>t}_{last,i}$, then the estimator can estimate $y_i(t)$ as:
\begin{equation}
y_i\left(t_{last,i}\right)-{\delta }_i\le \ y_i\left(t\right)\le y_i\left(t_{last,i}\right)+{\delta }_i
\end{equation}

The last received \textit{i}-th sensor data is used to compute the output $y_{computed,i}$ even if there is no sensor data transmission:
\begin{equation}
\label{eq:2}
y_{computed,i}\left(t\right)=y_i\left(t_{last,i}\right)=C_ix\left(t\right)+v_i\left(t\right)+{\Delta }_i\left(t,t_{last,i}\right)
\end{equation}
where ${\mathrm{\Delta }}_i\left(t,t_{last,i}\right)\mathrm{=}y_i\left(t_{last,i}\right)\mathrm{-}y_i\left(t\right)$ and: 

\begin{equation}
\label{eq:3}
\left|{\Delta }_i\left(t,t_{last,i}\right)\right|\le {\delta }_i
\end{equation}

In (\ref{eq:2}), the measurement noise increases from $v_i\left(t\right)$ to $v_i\left(t\right)+{\Delta }_i\left(t,t_{last,i}\right)$. If ${\Delta }_i\left(t,t_{last,i}\right)$ is assumed to have the uniform distribution with (\ref{eq:3}), then the variance of ${\Delta }_i\left(t,t_{last,i}\right)$ is $\frac{{\delta }^{\mathrm{2}}_i}{\mathrm{3}}$, which is added to the \textit{measurement noise covariance} in standard Kalman filter R$\left(i,i\right)$ when (\ref{eq:2}) applies.

\textbf{Improved Kalman Measurement Update Algorithm:} An algorithm is proposed here to appropriately improve the \textit{measurement update} part of the standard Kalman filter algorithm, which is adapted to the SoD event-generation condition by increasing the measurement noise covariance $\overline{R}_k$:
 \begin{enumerate}
 \item  Initialization set
 \begin{IEEEeqnarray}{ll}
 \hat{x}^-(0),{P}^-_0 \IEEEnonumber \\ 
 y_{last}=C\hat{x}^-\left(0\right)
 \end{IEEEeqnarray}
 \item  \textbf{Measurement update
 \begin{equation}
 \overline{R}_k=R
 \end{equation}
 if \textit{i}-th measurement data are received
 \begin{equation}
 \hat{y}_{last,i}=y_i\left(kT\right)
 \end{equation}
else
\begin{equation}
\overline{R}_k\left(i,i\right)=\overline{R}_k\left(i,i\right)+\frac{{\delta }^2_i}{3}
\end{equation}
end if
\begin{IEEEeqnarray}{lll}
K_k={P}^-_kC'(C{P}^-_kC'+\overline{R}_k)^{-1}\IEEEnonumber \\ 
\hat{x}\left(kT\right)=\hat{x}^-\left(kT\right)+K_k(\hat{y}_{last}-C\hat{x}^-(kT))\IEEEnonumber\\
P_k{=(I-K_kC)P}^-_k
\end{IEEEeqnarray}
}
\item  Project ahead
\begin{IEEEeqnarray}{lll}
\hat{x}^-\left((k+1)T\right)=\exp{\left(AT\right)}\hat{x}\left(kT\right)\IEEEnonumber\\
{P}^-_{k+1}=\exp{\left(AT\right)} P_k\exp{\left(A'T\right)}+Q_d
\end{IEEEeqnarray}
 \end{enumerate}
where $Q_d$ is the process noise covariance for the discretized dynamic system; $y_{last}$ is defined as (\ref{eq:jhjf}):
\begin{equation}\label{eq:jhjf}
y_{last}=[y_{last,1},y_{last,2},\dots ,y_{last,p}]'
\end{equation}

The presented event-triggered Kalman filter has been developed to implement the distributed controller and estimator as an NCS. It should be noted that in the proposed event-triggered observer, convergence is obtained by using the Kalman optimal observer. However, choosing the lower values of ${\delta }_i$ would result in the considerable reduction in the convergence time \cite{Li2016}. The controllers only receive updates from their neighbour controllers which is reflected in the $\textbf{L}$ matrix of the transfer function that has been realized. Distributed average consensus is then achieved for each estimator based on the number of neighbour controllers. Also, the higher the number of adjacent controllers are, the faster the estimator would converge \cite{Postoyan2015}.

\section{Developed Experimental Prototype for Situational Awareness}\label{prototype}

\begin{figure}[!t]
\centering
\includegraphics[width=3.5in]{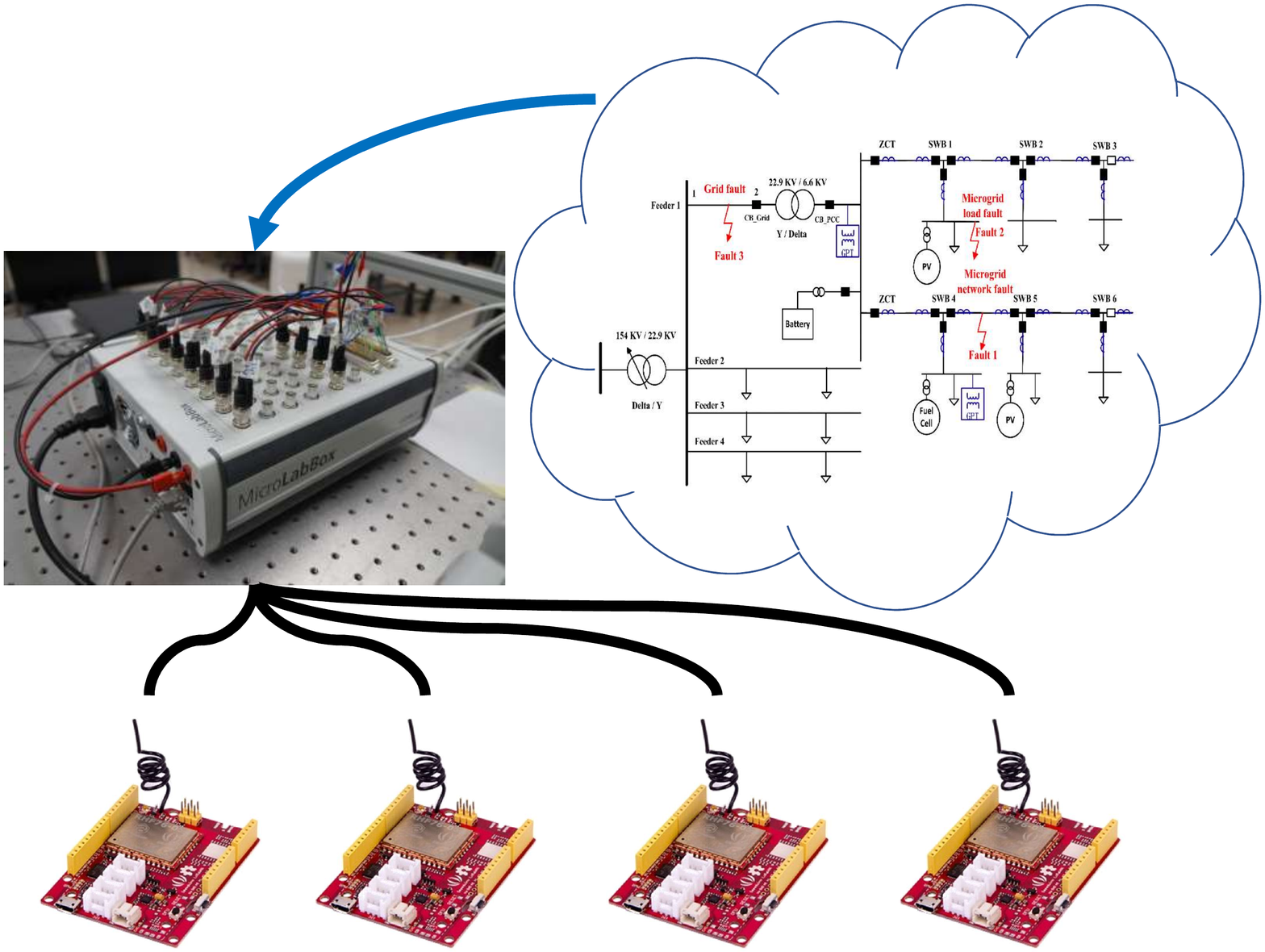}
\caption{Developed setup with the connected LoRaWAN sensor nodes.}
\label{hardware}
\end{figure}

The developed IoT setup consists of several nodes supporting long range wide area network (LoRaWAN) communication protocol from Seeeduino\textsuperscript{\textregistered} (Seeeduino LoRaWAN) and a real-time microgrid simulator from dSPACE\textsuperscript{\textregistered} (Microlabbox DS1202). The nodes are connected to the real-time simulator via the BNC connectors that can be both Analog Outputs and Analog Inputs. The schematic of the setup is shown in \figurename ~\ref{hardware}. The real-time simulator allows the testing of different microgrid operation scenarios with only changing the simulation configuration in the MATLAB software.

\begin{figure}[!t]
\centering
\includegraphics[width=3in,height=2in]{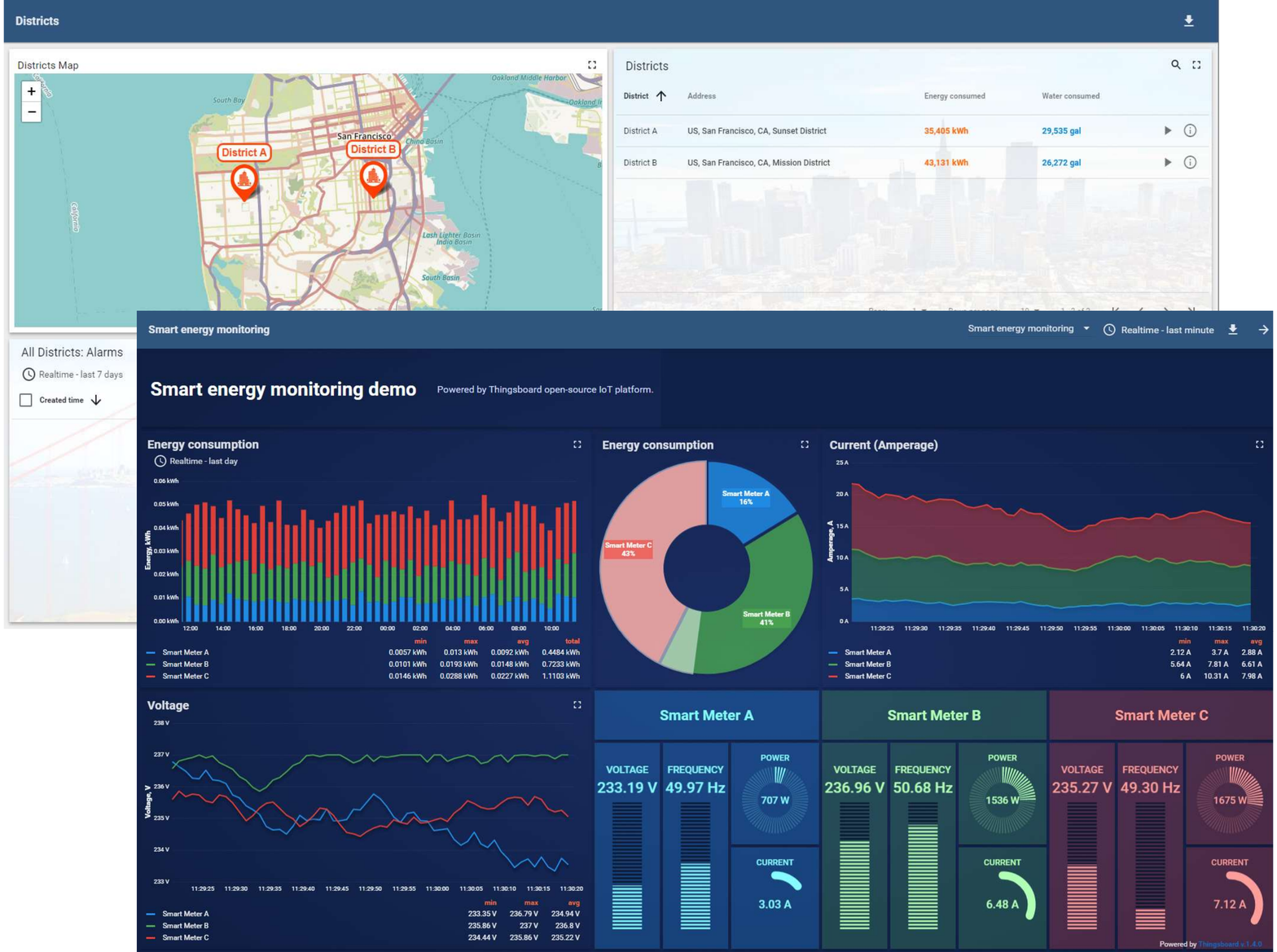}
\caption{Developed web-based dashboard using the Thingsboard IoT platform.}
\label{dashboard}
\end{figure}

Since microgrids will be installed in private urban or rural areas, the monitoring software should be accessible easily by the operators, and also a well-designed human machine interface (HMI) is essential, in order to achieve the adequate SA. In this regard, the web-based dashboards are suitable for this purpose, as they can be remotely accessed. In this work, the Thingsboard\textsuperscript{\textregistered} open-source software is used as the operator dashboard. Thingsboard is a web-based dashboard designer written in Java which provides different widgets to visualise the values received from the developed nodes. \figurename ~\ref{dashboard} shows the dashboard interface developed using the HTML5, CSS, and Javascript programming languages.

\begin{figure}[!t]
\centering
\includegraphics[width=3.5in,height=2.5in]{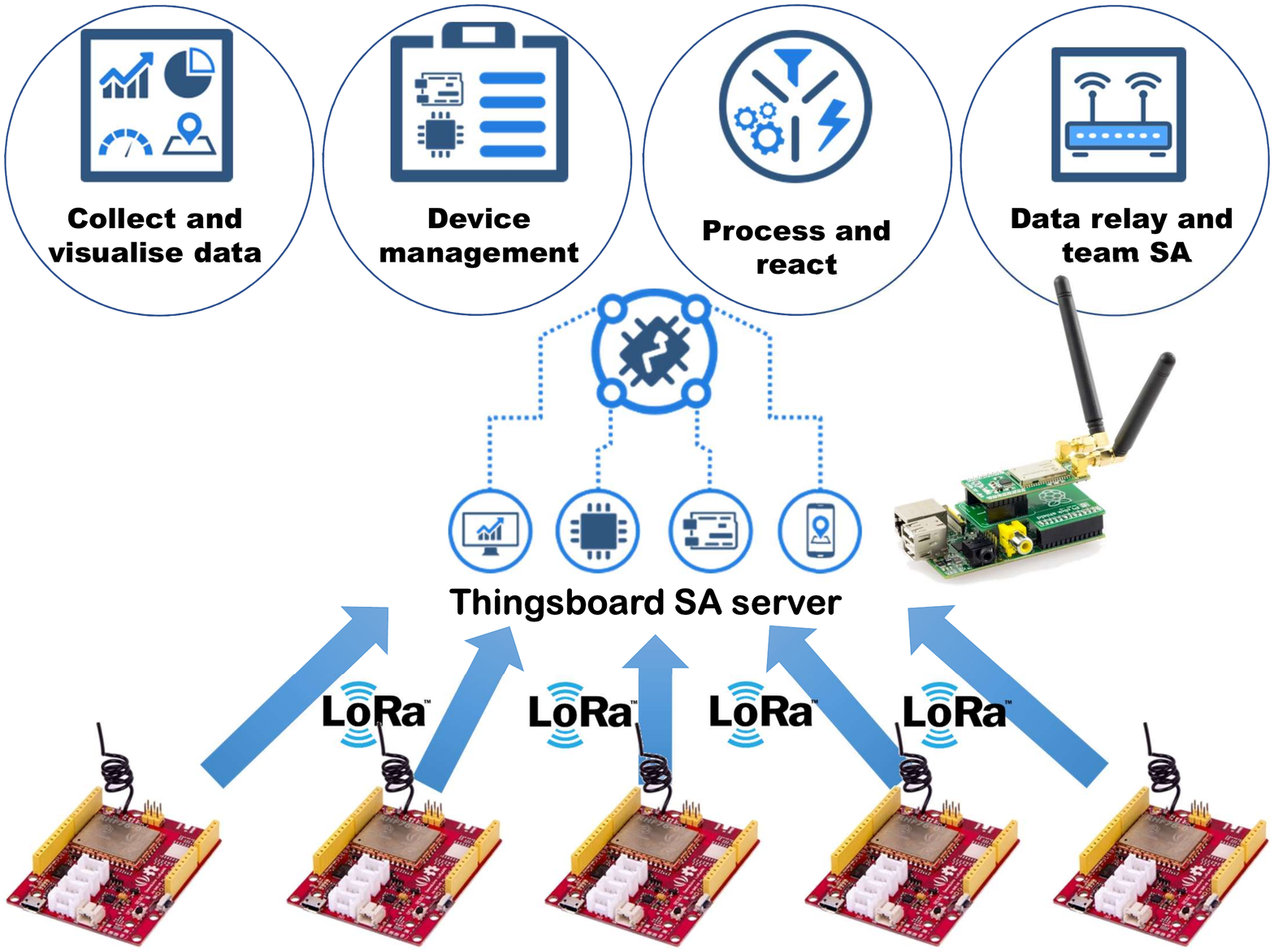}
\caption{Situational awareness-centric platform based on the IoT protocols.}
\label{setup}
\end{figure}

The LoRaWAN protocol necessitates a gateway to be employed for the data collection and distribution. In this setup, a Raspberry Pi with proper communuication modules for the gateway operation is used. This gateway converts the received data from LoRaWAN nodes and transforms them into MQTT payloads which are transmitted to the MQTT broker. Thingsboard IoT software provides the MQTT broker which in this work is employed for data processing and archiving. The data collection architecture for the situational awareness-centric microgrid platform based on the IoT protocols is shown in \figurename ~\ref{setup}. By using the mentioned protocols and devices, the cost of monitoring of smart grid is greatly reduced for realising the adequate SA. The upper layers in SA usually need different types of data, in order to analyse the current state of microgrid. The developed hardware setup is comparably more affordable than the existing monitoring devices, which makes it an ideal choice for the big data collection and processing in smart grid.

The software stack developed for this device, fully supports the Arduino\textsuperscript{\textregistered} integrated development environment (IDE). Many libraries are developed for the Arduino that can be used seamlessly in this device. In addition, the battery life is extended due to the event-based communication. Hence, lower rating batteries can be used that leads to the cost reduction.

\section{Results}\label{results}
\begin{figure}[h]
\centering
\includegraphics[width=1.5in]{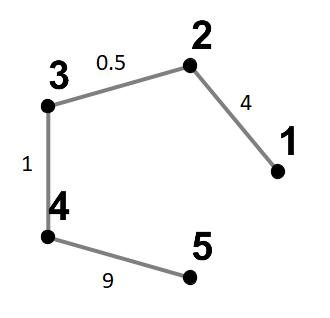}
\caption{Communication graph between the estimation agents.}
\label{graph}
\end{figure}

In order to evaluate the proposed state estimation approach, in this section, an example is given based on the developed multi-agent system with a distributed average consensus control. Consider a first-order multi-agent system defined as:
\begin{equation}
\dot{x_i} = u_i
\end{equation}
The state of the controller agents are denoted by the vector $x(t)=[x_1(t)^T, x_2(t)^T, \dots, x_5(t)^T]$, and the initial conditions are set as $x^0= [52, 44, 47, 48, 49]$ with the average of 48 volts. The topology graph shown in \figurename ~\ref{graph} is connected, i.e., is the requirement for the stability.

\begin{figure}[!h]
    \centering
	\subfloat{\includegraphics[width=4.3cm,height=4.3cm]{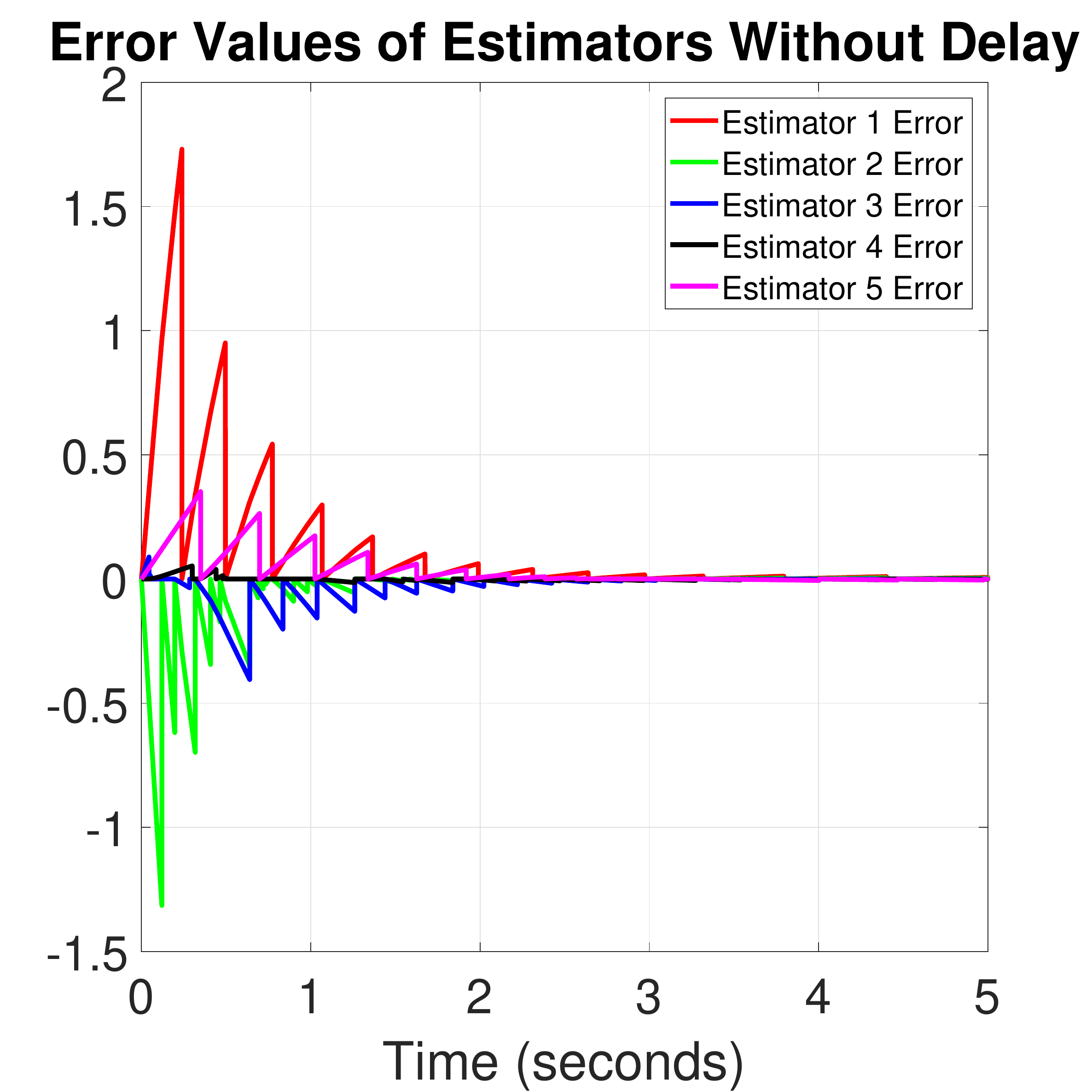}}
    \enspace
	\subfloat{\includegraphics[width=4.3cm,height=4.3cm]{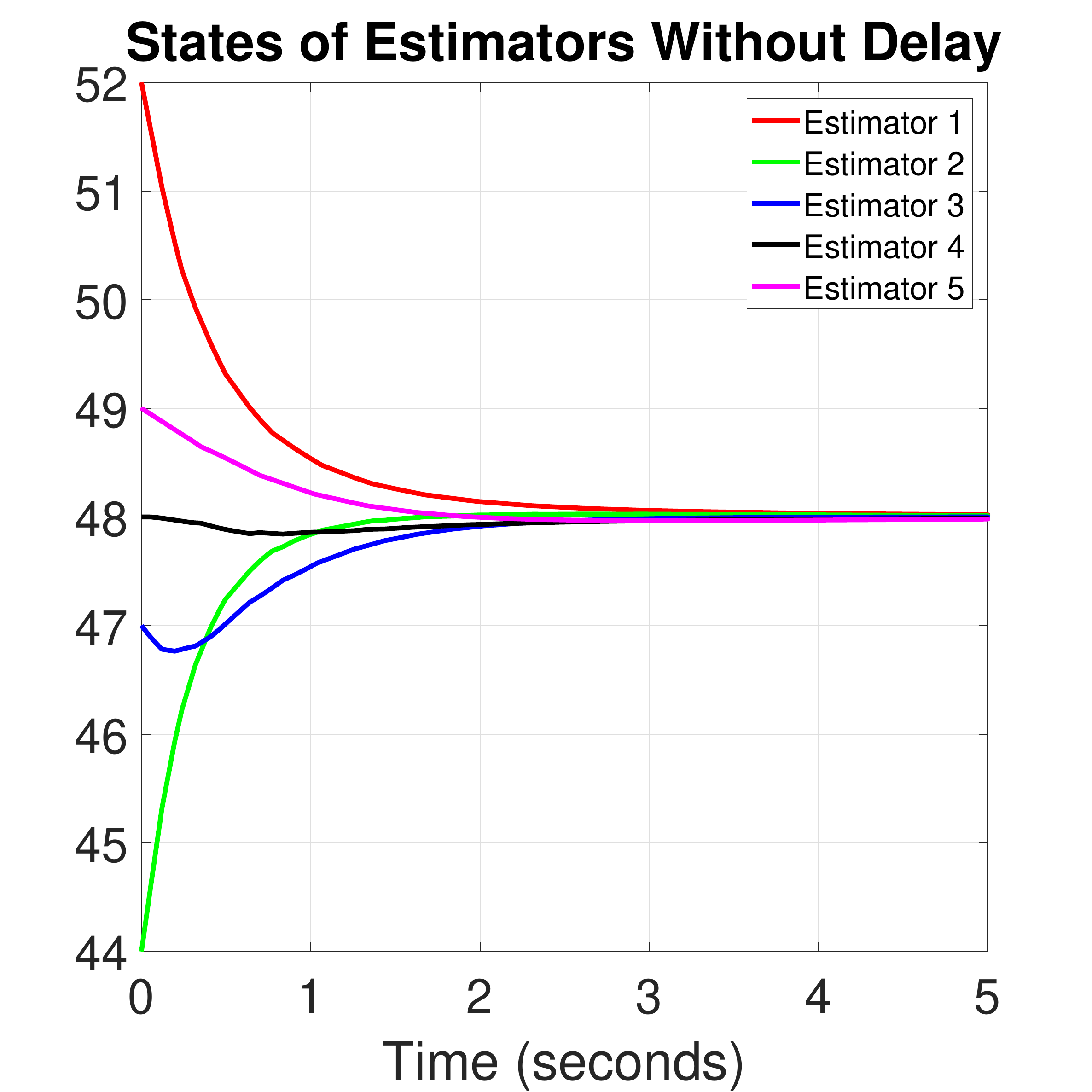}}
    \caption{States and errors of the distributed estimator agents \textbf{without} delay.}
    \label{fig:nodelay}
\end{figure}

\begin{figure}[!h]
    \centering
    \subfloat{\includegraphics[width=4.3cm,height=4.3cm]{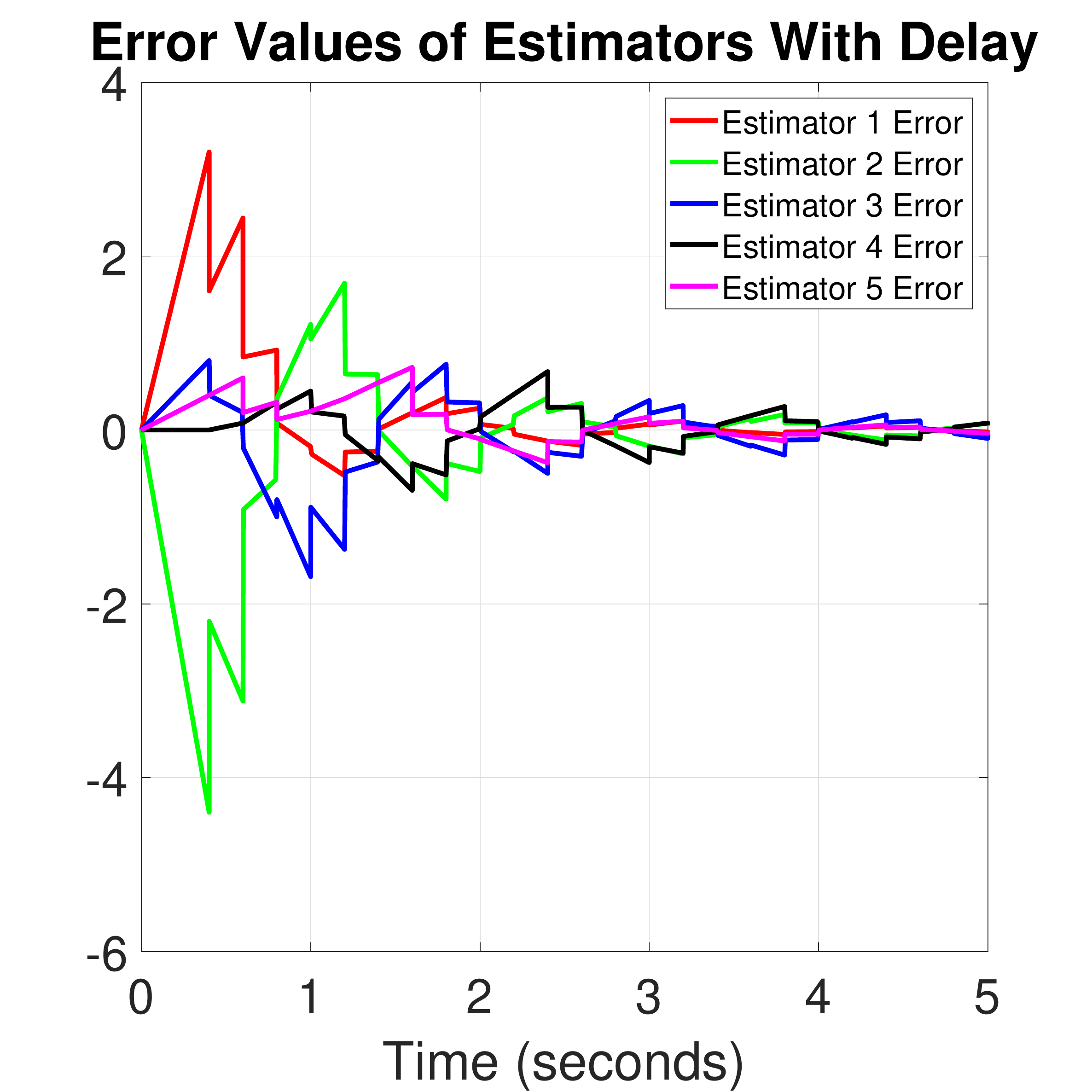}}
    \enspace
    \subfloat{\includegraphics[width=4.3cm,height=4.3cm]{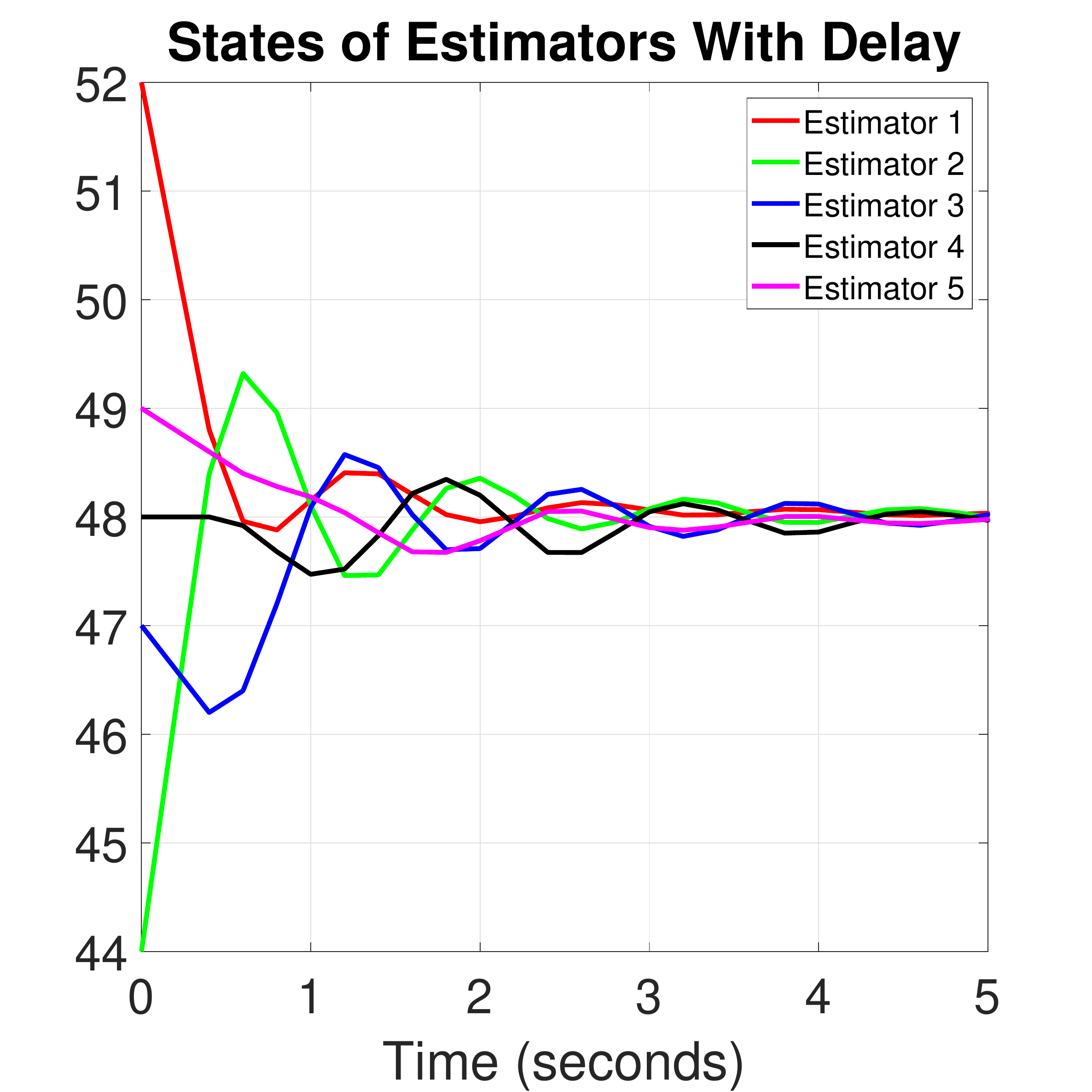}}    
    \caption{States and errors of the distributed estimator agents \textbf{with} delay.}
    \label{fig:delay}
\end{figure}

The two simulations are given; a simulation without delay (i.e., \figurename ~\ref{fig:nodelay}) and one with delay of 150ms (i.e., \figurename ~\ref{fig:delay}), in the event transmission, when the event-triggering condition is violated.

In both these figures, the state and measurement error values of the DG controller agents are shown. The measurement error values' profile illustrates a fluttering behavior. It presents when the agents are triggered at an event instant, the norm of the measurement error $e_i(t)$ is set to zero, due to the update of the state, when there is no delay. In contrast, as shown in \figurename ~\ref{fig:nodelay}, the error decay rate is higher comparing to the one with delay, i.e., in \figurename ~\ref{fig:delay}. It can be seen the event-triggered control strategy performs well in an environment with switching topologies.

\begin{table}[h]
\renewcommand{\arraystretch}{1.3}
\caption{Parameters of the Event-Triggered Kalman Estimator.}
\label{tab:kalmanparamstbl}
\centering
\begin{tabular}{|c|c|}
\hline
$\delta_i  (Voltage)$ &  0.1 V \\
\hline
$\delta_i (Energy)$ &  0.01 p.u. \\
\hline
$Q$ & 0 \\
\hline
$R$ & 1 \\
\hline
$T$ & 1 Second \\
\hline
\end{tabular}
\end{table}

\section{Conclusion}\label{conc}
This paper presents a data collection architecture and an event-triggered estimation strategy for situational awareness in microgrids. A setup has been developed which can provide enormous data collection capabilities from smart meters, in order to realise an adequate SA level in microgrids. It is shown by using the developed estimation strategy, an adequate level of SA can be achieved with a minimum installation and communication cost to have an accurate average state estimation of the microgrid.

\ifCLASSOPTIONcaptionsoff
  \newpage
\fi

\bibliographystyle{myIEEEtranbibstyle}

\bibliography{IEEEabrv,ref}

\begin{thebibliography}{10}
\providecommand{\url}[1]{#1}
\csname url@samestyle\endcsname
\providecommand{\newblock}{\relax}
\providecommand{\bibinfo}[2]{#2}
\providecommand{\BIBentrySTDinterwordspacing}{\spaceskip=0pt\relax}
\providecommand{\BIBentryALTinterwordstretchfactor}{4}
\providecommand{\BIBentryALTinterwordspacing}{\spaceskip=\fontdimen2\font plus
\BIBentryALTinterwordstretchfactor\fontdimen3\font minus
  \fontdimen4\font\relax}
\providecommand{\BIBforeignlanguage}[2]{{%
\expandafter\ifx\csname l@#1\endcsname\relax
\typeout{** WARNING: IEEEtran.bst: No hyphenation pattern has been}%
\typeout{** loaded for the language `#1'. Using the pattern for}%
\typeout{** the default language instead.}%
\else
\language=\csname l@#1\endcsname
\fi
#2}}
\providecommand{\BIBdecl}{\relax}
\BIBdecl

\bibitem{Montazeri2013}
M.~Montazeri, A.~Alavi, H.~{Rahmat Jou}, J.~{Mehr Ardestani}, and H.~{Jadali
  Pour}, ``{Real time substation distributed control system simulator
  development based on IEC 61850 standard for a sample substation: Case study:
  Sheikh Bahayi substation 400/230/63KV},'' in \emph{Smart Grid Conf. (SGC)},
  Dec. 2013, pp. 108--112.

\bibitem{Anderson2017}
A.~Anderson, P.~Loomba, I.~Orajaka, J.~Numfor, S.~Saha, S.~Janko, N.~Johnson,
  R.~Podmore, and R.~Larsen, ``{Empowering smart communities: Electrification,
  education, and sustainable entrepreneurship in IEEE smart village
  initiatives},'' \emph{IEEE Electrification Magazine}, vol.~5, no.~2, pp.
  6--16, Jun. 2017.

\bibitem{Alavi2018a}
S.~A. Alavi, K.~Mehran, Y.~Hao, A.~Rahimian, H.~Mirsaeedi, and V.~Vahidinasab,
  ``{A distributed event-triggered control strategy for DC microgrids based on
  publish-subscribe model over industrial wireless sensor networks},''
  \emph{IEEE Transactions on Smart Grid}, 2018.

\bibitem{Sharifzadeh2017a}
M.~Sharifzadeh, F.~Separi, and M.~Heydari, ``{Planning of autonomous smart
  micro grid for electrification of remote villages in MEDC},'' \emph{CIRED -
  Open Access Proceedings Journal}, vol. 2017, no.~1, pp. 2493--2495, Oct.
  2017.

\bibitem{Alavi2016}
A.~Alavi, M.~Javadipour, and A.~A. Afzalian, ``{An optimal event-triggered
  tracking control for battery-based wireless sensor networks},'' in
  \emph{Smart Grids Conf. (SGC)}, Dec. 2017, pp. 42--47.

\bibitem{Basu2016}
C.~Basu, M.~Padmanaban, S.~Guillon, L.~Cauchon, M.~{De Montigny}, and I.~Kamwa,
  ``{Situational awareness for the electrical power grid},'' \emph{IBM Journal
  of Research and Development}, vol.~60, no.~1, pp. 10:1--10:11, Jan. 2016.

\bibitem{Diao2010}
R.~Diao, V.~Vittal, and N.~Logic, ``{Design of a real-time security assessment
  tool for situational awareness enhancement in modern power systems},''
  \emph{IEEE Transactions on Power Systems}, vol.~25, no.~2, pp. 957--965, May
  2010.

\bibitem{Monajemi2017}
T.~Monajemi, A.~Rahimian, and K.~Mehran, ``{Energy management using a
  situational awareness-centric ad-hoc network in a home environment},'' in
  \emph{2nd EAI Int. Conf. Smart Grid Inspired Future Technologies
  (SmartGIFT)}, Mar. 2017, pp. 15--24.

\bibitem{Wang2014}
W.~Wang, L.~He, P.~Markham, H.~Qi, Y.~Liu, Q.~C. Cao, and L.~M. Tolbert,
  ``{Multiple event detection and recognition through sparse unmixing for
  high-resolution situational awareness in power grid},'' \emph{IEEE
  Transactions on Smart Grid}, vol.~5, no.~4, pp. 1654--1664, Jul. 2014.

\bibitem{Alavi2018}
S.~A. Alavi, A.~Rahimian, K.~Mehran, and J.~Mehr~Ardestani, ``{An IoT-based
  data collection platform for situational awareness-centric microgrids},'' in
  \emph{IEEE Canadian Conf. Electrical {\&} Computer Eng. (CCECE)}, May 2018.

\bibitem{Spanos2005DynamicCF}
D.~P. Spanos, R.~Olfati-saber, and R.~M. Murray, ``{Dynamic consensus for
  mobile networks},'' in \emph{16th Int. Fed. Aut. Control (IFAC)}, 2005, pp.
  1--6.

\bibitem{Morstyn2016}
T.~Morstyn, B.~Hredzak, G.~D. Demetriades, and V.~G. Agelidis, ``{Unified
  distributed control for DC microgrid operating modes},'' \emph{IEEE
  Transactions on Power Systems}, vol.~31, no.~1, pp. 802--812, Jan. 2016.

\bibitem{DePersis2018}
C.~{De Persis}, E.~R. Weitenberg, and F.~D{\"{o}}rfler, ``{A power consensus
  algorithm for DC microgrids},'' \emph{Automatica}, vol.~89, pp. 364--375,
  Mar. 2018.

\bibitem{Li2016}
W.~Li, J.~Du, and Y.~Jia, ``{Event-triggered Kalman consensus filter over
  sensor networks},'' \emph{IET Control Theory {\&} Applications}, vol.~10,
  no.~1, pp. 103--110, Jan. 2016.

\bibitem{Postoyan2015}
R.~Postoyan, P.~Tabuada, D.~Ne{\v{s}}i{\'{c}}, and A.~Anta, ``{A framework for
  the event-triggered stabilization of nonlinear systems},'' \emph{IEEE
  Transactions on Automatic Control}, vol.~60, no.~4, pp. 982--996, Apr. 2015.

\end{thebibliography}

\end{document}